\documentclass[slac_one]{revtex4}
\usepackage{graphicx}
\usepackage{fancyhdr}
\begin{document}
\def\be{\begin{equation}}
\def\ee{\end{equation}}

\title{Large $N$ lattice gauge theory} 

%

\author{Rajamani Narayanan\footnote{speaker}}
\affiliation{Department of Physics,
Florida International University, Miami, FL 33199}
\author{Herbert Neuberger}
\affiliation{
        Department of Physics and Astronomy, Rutgers University,
Piscataway, NJ 08855}

\begin{abstract}
Wilson loops in large N gauge theory exhibit a weak to strong coupling
transition as the loop is dilated. A multiplicative matrix model
captures the universal behavior associated with this transition. A
universal scaling function is obtained in a double scaling
limit. Numerical studies show that both large N QCD in three
dimensions and the SU(N) principal chiral model in two dimensions are
in the same universality class.
\end{abstract}

\maketitle

\thispagestyle{fancy}

Large $N$ gauge theories may provide a bridge between
QCD and string theory. We hope that this would produce a 
way to make a calculational 
connection
between some weak and strong coupling phenomena
of continuum QCD. Recent 
progress~\cite{Narayanan:2006rf,Narayanan:2007dv,Narayanan:2008he}
 in this direction is the
main subject of this talk.

We will consider large $N$ gauge theories on a 
$d=2,3,4$ dimensional continuum torus. It is sufficient for
our discussion to consider a symmetric torus of length
$l$ in all four directions. We will consider the
't Hooft limit~\cite{'tHooft:1973jz}
 with a finite number of fundamental fermion
flavors. There is no back reaction from the fermions
on the gauge field dynamics.

In all three dimensions,
there is a critical size $l_1$ such that continuum
reduction~\cite{Narayanan:2003fc}
 holds for $l>l_1$ and there are no finite
volume effects. The critical size, $l_1$, is equal
to $T_c^{-1}$ where $T_c$ is the deconfining temperature.
Two dimensions
is a special case since $l_1=0$ and large $N$ QCD is always
in the confined phase and there is no dependence on the
temperature. Therefore, continuum
Eguchi-Kawai reduction~\cite{Eguchi:1982nm}
holds in two dimensions~\cite{Bhanot:1982sh}.
We will focus on the confined phase of large $N$ QCD in this
talk. 

Let us first make a remark about the Gross-Witten
transition~\cite{Gross:1980he} 
in two dimensional large $N$ lattice gauge
theory. This
transition has analogs in large $N$ lattice gauge theories
in three and four dimensions. The 
eigenvalue distribution of the plaquette
operator opens up a gap around $-1$ at the transition point
enabling the presence of topologically disconnected
classes of gauge fields in the continuum limit~\cite{Kiskis:2002gr}.
These transitions do not have a continuum analog.
We will assume that we are on the physical side
of this transition when discussing large $N$ lattice
gauge theories.

The stage is now set to discuss the transition from weak 
coupling to strong coupling in the confined phase
of large $N$ QCD. 
We will use 
the Wilson loop operator, $W$, a
unitary operator for SU(N) gauge theories, 
to probe the transition. 
Large (area) Wilson loops are non-perturbative and
correspond to strong coupling.
Small (area) Wilson loops are perturbative and
correspond to weak coupling.

The probe is defined as
\be
{\cal W}_N(z,b,L)=\langle \det (z-W) \rangle
\ee
and is the characteristic polynomial associated with the
operator. 
$W$ is the Wilson loop operator, 
$z$ is a complex number, 
$N$ is the number of colors,
$b=\frac{1}{g^2N}$ is the lattice gauge coupling and
$L$ is the linear size of the square loop.
$\langle\cdots\rangle$ is the average over all gauge fields
with the standard gauge action.

The eigenvalues of $W$ are gauge invariant and so is the
characteristic polynomial. The eigenvalues lie on the
unit circle and all of them will be close to unity for
small loops. The eigenvalues will spread uniformly
over the unit circle for large loops. The characteristic
polynomial exhibits a transition 
at $N=\infty$ when $L\to L_c(b)$. 
This is a physical transition since $L_c(b)$
will scale properly with the coupling, $b$, as one
approaches the continuum limit. 

The eigenvalue distribution of the Wilson loop
operator in two dimensional
QCD was shown to undergo the above transition by
Durhuus and Olesen~\cite{Durhuus:1980nb}
and further relevant discussions can be found 
in~\cite{Douglas:1993iia,Bassetto:1999dg}.
The scaling function in the double scaling limit was
derived for two dimensional large $N$ QCD
in~\cite{Narayanan:2007dv}.
Two dimensional gauge theory on an infinite lattice can
be gauge fixed so that the only variables are the individual
plaquettes and these will be independently and identically
distributed. The Wilson loop operator, $W$, can be
written as
$W=\prod_{j=1}^n U_j$ where the $U_j$'s are the transporters
around the individual plaquettes that make up the loop
and $n=L^2$ is equal to the area of the loop.
The measure associated with any $U_j$ can be set to
$P(U_j) = {\cal N} e^{-\frac{N}{2} {\rm Tr\ } H_j^2}$
where $U_j = e^{i\epsilon H_j}$ and $\epsilon$ plays
the role of gauge coupling.
The dimensionless area is given by $t=\epsilon^2 n$ 
which
is kept fixed as one takes the
 limit $n\to\infty$ and $\epsilon\to 0$.
This is the multiplicative matrix model of~\cite{janik}.
In the continuum limit,
the parameters $b$ and $L$ get replaced by one parameter,
which is denoted by
$t$ in the model, and the characteristic polynomial
becomes
\begin{eqnarray}
{\cal W}_N(z,b,L) \to Q_N(z,t)
 &=&\cases{
\sqrt{\frac{N\tau}{2\pi}}
\int_{-\infty}^\infty d\nu
e^{-\frac{N}{2}\tau\nu^2} 
\left[z-e^{-\tau\nu
-\frac{\tau}{2}
}
\right]^N &  $SU(N)$ \cr
\sqrt{\frac{Nt}{2\pi}}
\int_{-\infty}^\infty d\nu
e^{-\frac{N}{2}t\nu^2} 
\left[z-e^{-t\nu
-\frac{\tau}{2}
}
\right]^N &  $U(N)$ \cr
}\cr
 &=&\cases{
\sum_{k=0}^N 
\pmatrix{
N\cr k\cr
}
z^{N-k} (-1)^k e^{-\frac{\tau k(N-k)}{2N}} &
 $SU(N)$ \cr
\sum_{k=0}^N 
\pmatrix{
N\cr k\cr}
z^{N-k} (-1)^k e^{-\frac{t k(N+1-k)}{2N}} &
$U(N)$ \cr
};\ \ \ \ \tau=t\left(1+\frac{1}{N}\right)
\label{qnint}
\end{eqnarray}
\be
\ee

A rearrangement of $Q_N(z,t)$ for $SU(N)$,
\be
Z_N(z,t)=
Q_N(z,t)(-1)^N e^{\frac{(N-1)\tau}{8}} (-z)^{-\frac{N}{2}}=
\sum_{\sigma_1,\sigma_2,...\sigma_N =\pm\frac{1}{2}} 
e^{\ln(-z)\sum_i \sigma_i}
e^{\frac{\tau}{N} \sum_{i > j} \sigma_i \sigma_j},
\ee
puts it in the form of 
the partition function for a spin model with
a ferromagnetic interaction for positive $\tau$ with
$ln(-z)$ as a complex external magnetic field.
Therefore, the conditions for Lee-Yang theorem~\cite{Lee:1952ig}
 are fulfilled
and all roots of $Q_N(z,t)$ lie on the unit circle for SU(N).

The transition from weak coupling to strong coupling can
be intuitively seen using the characteristic polynomial,
$Q_N(z,t)$.
In the weak coupling (small area) limit we have $t=0$ and
$Q_N(z,t)=(z-1)^N$. Therefore, 
all roots are at $z=1$ on the unit circle.
In the strong coupling (large area) limit we have $t=\infty$
and
$Q_N(z,t)=z^N+(-1)^N$.
Therefore, all roots are uniformly distributed on the unit circle.

$Q_N(z,t)$ is analytic in $z$ for all $t$ at finite $N$.
But, this is not the case as $N\to\infty$ and 
leads to a transition from weak to strong coupling in the
$N\to\infty$ limit.

There is a critical area, $t=4$, such that
the distribution of zeros of $Q_\infty(z,t)$ 
on the unit circle has a gap around $z=-1$
for $t < 4$ and has no gap for $t > 4$~\cite{janik,Durhuus:1980nb}.
To see this, we note that
the integral representation (\ref{qnint}) 
is dominated by the saddle point, $\nu=\lambda(t,z)$, given by
\be
\lambda=\lambda(t,z)=\frac{1}{ze^{t(\lambda+\frac{1}{2})}-1}
\ee
With
$z=e^{i\theta}$
and $w=2\lambda+1$, 
$\rho(\theta)=-\frac{1}{4\pi}{\bf Re}\ w$
gives the distribution of the eigenvalues of $W$ on the unit circle.

The saddle point equation at $z=-1$ is
\be
w=\tanh \frac{t}{4}w
\ee
showing that $w$ admits non-zero real solutions for $t>4$.

As $N\to\infty$, one can define a scaling region around
$t=4$ and $z=-1$ by
\be
t=\frac{4}{1+\frac{\alpha}{\sqrt{3N}}};\ \ \ \
z=-e^{\left(\frac{4}{3N}\right)^{\frac{3}{4}}\xi}
\ee
$\alpha$ and $\xi$ are the scaling variables that
blow up the region near $t=4$ and $z=-1$.
We can show that
\be
\lim_{N\rightarrow\infty}
\left(\frac{4N}{3}\right)^{\frac{1}{4}}
(-1)^N e^{\frac{(N-1)\tau}{8}} (-z)^{-\frac{N}{2}}
Q_N(z,t)
=
\int_{-\infty}^{\infty} du e^{-u^4-\alpha u^2+\xi u }
\equiv \zeta(\xi,\alpha)
\ee
which is the scaling function in the double scaling limit
associated with the characteristic polynomial.

This scaling behavior is expected to be universal and
was numerically tested 
using the lattice formulation in three dimensional
large $N$ QCD~\cite{Narayanan:2007dv}
and in the two dimensional SU(N)
principal chiral model~\cite{Narayanan:2008he}. 
The precise statement of the
continuum
large $N$ universality
hypothesis is as follows.
Let ${\cal C}$ be a closed 
non-intersecting loop: $x_\mu(s), s\in[0,1]$.
Let ${\cal C}(m)$ be a whole family of loops obtained by
dilation: $x_\mu(s,m)=\frac{1}{m} x_\mu 
(s)$,with $m > 0.$ 
Let $W(m,{\cal C}(*))=
W({\cal C}(m))$ be the family of operators associated
with the family of loops denoted by ${\cal C}(*)$
where $m$ labels one member in the family.
Define
\be
O_N (y,m,{\cal C}(*))=\langle \det (e^{\frac{y}{2}}+e^{-\frac{y}{2}} 
W(m,{\cal C}(*))\rangle
\ee
Then our hypothesis is
\be
\lim_{N\rightarrow\infty} {\cal N}(N,b,{\cal C}(*))
O_N\left(y=
\left(\frac{4}{3N^3}\right)^{\frac{1}{4}}\frac{\xi}{a_1({\cal C}(*))},
m=m_c\left [1+\frac{\alpha}{\sqrt{3N}a_2({\cal C}(*))}\right ]
\right) = 
\zeta(\xi,\alpha)
\ee

Even though the hypothesis has not been explicitly tested in
four dimensional large $N$ QCD, there is enough numerical
evidence in suppport of the hypothesis~\cite{Narayanan:2006rf}. 
Assuming the universal behavior, one can address the question
of matching the universal data in the transition region to
the perturbative side.

\acknowledgments

R.N. acknowledge partial support by the NSF under grant number
PHY-055375 at Florida International University.  
H. N. acknowledges partial support by the DOE, grant \#
DE-FG02-01ER41165, and the SAS of Rutgers University.


\begin{thebibliography}{99}
\bibitem{Narayanan:2006rf}
  R.~Narayanan and H.~Neuberger,
  JHEP {\bf 0603}, 064 (2006)
  [arXiv:hep-th/0601210].
\bibitem{Narayanan:2007dv}
  R.~Narayanan and H.~Neuberger,
  JHEP {\bf 0712}, 066 (2007)
  [arXiv:0711.4551 [hep-th]].
\bibitem{Narayanan:2008he}
  R.~Narayanan, H.~Neuberger and E.~Vicari,
  JHEP {\bf 0804}, 094 (2008)
  [arXiv:0803.3833 [hep-th]].
\bibitem{'tHooft:1973jz}
  G.~'t Hooft,
  ``A PLANAR DIAGRAM THEORY FOR STRONG INTERACTIONS,''
  Nucl.\ Phys.\  B {\bf 72}, 461 (1974).
\bibitem{Narayanan:2003fc}
  R.~Narayanan and H.~Neuberger,
  Phys.\ Rev.\ Lett.\  {\bf 91}, 081601 (2003)
  [arXiv:hep-lat/0303023].
\bibitem{Eguchi:1982nm}
  T.~Eguchi and H.~Kawai,
  ``Reduction Of Dynamical Degrees Of Freedom In The Large N Gauge Theory,''
  Phys.\ Rev.\ Lett.\  {\bf 48}, 1063 (1982).
\bibitem{Bhanot:1982sh}
  G.~Bhanot, U.~M.~Heller and H.~Neuberger,
  ``The Quenched Eguchi-Kawai Model,''
  Phys.\ Lett.\  B {\bf 113}, 47 (1982).
\bibitem{Gross:1980he}
  D.~J.~Gross and E.~Witten,
  ``Possible Third Order Phase Transition In The Large N Lattice Gauge
  Theory,''
  Phys.\ Rev.\  D {\bf 21}, 446 (1980).
\bibitem{Kiskis:2002gr}
  J.~Kiskis, R.~Narayanan and H.~Neuberger,
  Phys.\ Rev.\  D {\bf 66}, 025019 (2002)
  [arXiv:hep-lat/0203005].
\bibitem{Durhuus:1980nb}
  B.~Durhuus and P.~Olesen,
  Nucl.\ Phys.\  B {\bf 184}, 461 (1981).
\bibitem{Douglas:1993iia}
  M.~R.~Douglas and V.~A.~Kazakov,
  Phys.\ Lett.\  B {\bf 319}, 219 (1993)
  [arXiv:hep-th/9305047].
\bibitem{Bassetto:1999dg}
  A.~Bassetto, L.~Griguolo and F.~Vian,
  ``Instanton contributions to Wilson loops with general winding number in  two
  dimensions and the spectral density,''
  Nucl.\ Phys.\  B {\bf 559}, 563 (1999)
  [arXiv:hep-th/9906125].
\bibitem{janik} R. A. Janik, W. Wieczorek, J.\ Phys.\ A:\ Math.\ Gen.
{\bf 37}, 6521 (2004).
\bibitem{Lee:1952ig}
  T.~D.~Lee and C.~N.~Yang,
  Phys.\ Rev.\  {\bf 87}, 410 (1952).
\end{thebibliography}
\end{document}